# Tunable phonon cavity coupling in graphene membranes

R. De Alba[1,*], F. Massel[2], I. R. Storch[1], T. S. Abhilash[1], A. Hui[3], P. L. McEuen[1,4], H. G. Craighead[3] & J. M. Parpia[1]

[1]Department of Physics, Cornell University, Ithaca, New York 14853, USA
[2]Department of Physics, Nanoscience Center, University of Jyväskylä, F1-40014, Finland
[3]School of Applied & Engineering Physics, Cornell University, Ithaca, New York 14853, USA
[4]Kavli Institute at Cornell for Nanoscale Science, Ithaca, New York 14853, USA

**A major achievement of the past decade has been the realization of macroscopic quantum systems by exploiting interactions between optical cavities and mechanical resonators[1,2,3]. In these systems, phonons are coherently annihilated or created in exchange for photons. Similar phenomena have recently been observed through "phonon cavity" coupling – energy exchange between modes of a single system as mediated by intrinsic material nonlinearity[4,5]. To date, this has been demonstrated primarily for bulk crystalline, high-quality-factor ($Q > 10^5$) mechanical systems operated at cryogenic temperatures. Here we propose graphene as an ideal candidate for the study of such nonlinear mechanics. The large elastic modulus of this material and capability for spatial symmetry breaking via electrostatic forces is expected to generate a wealth of nonlinear phenomena[6], including tunable inter-modal coupling. We have fabricated circular graphene membranes and report strong phonon cavity effects at room temperature, despite the modest $Q$ (~100) of this system. We observe both amplification into parametric instability ("mechanical lasing") and cooling of Brownian motion in the fundamental mode through excitation of cavity sidebands. Furthermore, we characterize quenching of these parametric effects at large vibrational amplitudes, offering a window on the all-mechanical analogue of cavity optomechanics, where observation of such effects has**

*e-mail: rd366@cornell.edu

**proven elusive.**

Mechanical resonators composed of tensioned graphene membranes have been widely studied in recent years[7–14]; their low mass, $\rho_g \approx 0.75 \text{mg/m}^2$, electrical integrability, and strong optical interaction[11,15] make them rich and versatile systems studied largely for force and mass sensing. At room temperature their moderate $Q$'s, extreme frequency tunability, and low in-line resistance makes these structures promising as intermediate-frequency (1-50MHz) electromechanical elements, including passive filters and oscillators. At cryogenic temperatures ($T < 4\text{K}$) graphene is becoming an attractive system for the study quantum motion, as it exhibits both large zero point motion and drastically enhanced $Q$'s; progress towards this end has already been made[16–18], with coupling to on-chip microwave cavities and significant optomechanical cooling recently demonstrated. The nonlinearity studied here represents a complementary method for parametric control of these membranes based on the intrinsic interactions of vibrational modes. This all-mechanical effect can be utilized to enhance the $Q$ (and hence sensitivity) of graphene-based sensors, provide multi-mode readout through detection of a single mode[19], and ultimately enable information exchange between optically cooled quantum modes. Moreover, this coupling makes graphene viable as low-power, tunable, electromechanical frequency mixers.

The primary source of nonlinearity in graphene membranes is motion-induced tension modulation. Similar to mode coupling in other mechanical systems[4,20,21], one vibrational mode (here assumed to be the fundamental mode at frequency $\omega_1$) can be parametrically manipulated through its interaction with a second mode, which is deemed the phonon cavity (at $\omega_c$). Exciting the coupled system at the cavity's red sideband ($\omega_c - \omega_1$) results in energy flow from the fundamental to the cavity, whereas pumping the blue sideband ($\omega_c + \omega_1$) generates amplification of both the fundamental and cavity modes; these processes are depicted in Figure 1c. The efficiency of this inter-modal energy exchange is dictated by the coupling rate, $G = d\omega_c/dx_1$, where $x_1$ is the amplitude of motion at $\omega_1$. This coupling rate is reminiscent of cavity opto-mechanics, and an identical formalism can be used to derive the resulting equations of motion (supplementary material, section S1).

The advantages of graphene over other membrane materials (e.g. SiN) in generating this effect are two-fold: 1) For the coupling mechanism under consideration, $G$ increases linearly with the static membrane deflection, $x_0$. In graphene this value can be tuned electrostatically with a dc bias voltage. Moreover, because of its atomic thinness ($h \sim$ 0.3nm), graphene can withstand large out-of plane stretching. This is the result of an extremely low in-plane stiffness, $C = Eh/(1 - v^2)$, where $E$ and $v$ are the elastic modulus (1.0TPa for exfoliated[22] and 160GPa for CVD[23] graphene) and Poisson ratio, respectively. Previous studies of graphene have shown $x_0$ can exceed 3% of the suspended length without rupturing[24]. 2) Because the tension in graphene is highly tunable, the frequency spectrum can be adjusted to obtain 3-mode alignment, $\omega_c \pm \omega_1 \approx \omega_{sb}$. Here $\omega_{sb}$ signifies the resonance of a third mode, which overlaps the cavity sideband and enhances pumping by a factor of $Q_{sb}$; this arrangement is also depicted in Fig. 1c. Under these conditions, it is thus possible to generate large phonon cavity effects in the room temperature graphene system.

It should be noted that there are alternative inter-modal coupling mechanisms available for tensioned membranes – most notably, mutual coupling to a resonance of the surrounding substrate[21]. Such systems enable parametric membrane control in a manner qualitatively similar to the coupling studied here, but also necessitate the 3-mode alignment described above, which can be a challenge if the spectrum is not experimentally tunable. Moreover, a unique feature of the graphene system is the tunability of the coupling rate itself, $G \propto x_0$, which is present neither in the substrate-coupled case nor in standard optomechanics experiments.

We have fabricated circular graphene drums with diameters $d$ ranging from 5 to 20 µm; we report measurements of two drums – "Device 1" ($d = 8$ µm) and "Device 2" ($d = 20$ µm) – although the effects reported have been observed across a wide number of samples. A micrograph of Device 1 and diagram of the experimental setup are shown in Fig. 1a,b. Motion is driven electrostatically via an applied bias voltage $V_{dc} + v \sin \omega t$ and detected optically through laser interferometry[11]. Unlike previous generations of graphene resonators, our structures feature two independent back-gates, which enable efficient actuation of several mechanical modes. The gate-graphene separation is 1.7µm. Most measurements were performed with one gate grounded and a drive signal applied to the other, although other configurations (shown in Fig. 1a) can be used to favor either the fundamental mode or higher modes.

Device 1 has 6 modes that can be readily excited (Fig. 2a,b). The frequency dispersion of this spectrum with $V_{dc}$ is shown in Fig. 2a. Between $V_{dc} = 0 - 7.5$V, there is reasonable overlap between modes 2,6 and their respective sidebands (Fig. 2d); therefore this is where we expect the strongest phonon cavity effect. At $V_{dc} = 5V$, the graphene has natural frequencies and $Q$'s of: $\omega_1/2\pi = 8.6$ MHz, $\omega_2/2\pi = 12.4$ MHz, $\omega_6/2\pi = 21.0$ MHz, $Q_1 = \omega_1/\gamma_1 = 57$, $Q_2 = 48$, and $Q_6 = 37$.

The general Hamiltonian of our coupled system including the fundamental and cavity modes is

$$H = \sum_{n=1,c} \left( \frac{p_n^2}{2m_n} + \frac{1}{2}m\omega_n^2 x_n^2 + L_n x_n + S_n x_n^2 + T_n x_n^3 + F_n x_n^4 \right)$$

$$+ T_{1c} x_1 x_c^2 + T_{c1} x_c x_1^2 + F_{1c} x_1^2 x_c^2$$

Eq. 1

where $m$ is the membrane mass and $p_n$ is the momentum of mode $n$. The first line comprises the linear response of each mode, as well as self-nonlinearities that produce Duffing behavior. Terms in the second line reflect all possible interactions between the two modes up to fourth order. The coefficients $L, S, T,$ and $F$ include only tensioning effects, and their magnitudes are determined by the displacement profiles of the two modes; they are calculated explicitly for a circular membrane geometry (as well as a general geometry) in section S1 of the supplementary material. Mode profiles have been measured for Device 1 (Fig. 2b), and have stark differences from the expected Bessel functions; this can occur due to mass or tension inhomogeneity arising during fabrication, and can be mitigated by using a more sophisticated graphene clamping scheme[25]. The fourth-order coupling $F_{1c}$ generates an effective shift in $\omega_1^2$ (or $\omega_c^2$) proportional to $x_c^2$ (or $x_1^2$) but does not affect the damping of either mode. $T_{1c}$ and $T_{c1}$, on the other hand, enable parametric control of one mode based on sideband pumping of the other. For a perfect circular membrane, we have $T_{mn} = (C\alpha_n^2/\pi d^2) \cdot \int (\vec{\nabla} x_0 \cdot \vec{\nabla} \xi_m) dA$, where $\alpha_n$ is the Bessel-function zero of mode $n$, and $x_0(\vec{r})$, $\xi_m(\vec{r})$, are the static deflection profile and normalized profile of mode $m$. For the modes under consideration in Device 1, $T_{c1} \approx 0$ due to the symmetry of the cavity modes.

Within a linearized description of the phonon cavity coupling, the coupling rate is $G(\omega) = d\omega_c/dx_1 = T_{1c}/m\omega_c$. We probe the effects of mode coupling by applying two concurrent drive signals: a probe signal at frequency $\omega$, around $\omega_1$, and a pump signal at $\omega_p$. In terms of the cavity detuning $\Delta = \omega_p - \omega_c$ and the pumped vibration amplitude $x_p = x(\omega_p)$, the effective resonant frequency and damping of mode 1 are:

$$\omega_{1,\text{eff}} = \Omega_1 + \frac{2G^2|x_p|^2 \Delta[\gamma_c^2/4 - \omega^2 + \Delta^2]}{[\gamma_c^2/4 + (\omega - \Delta)^2][\gamma_c^2/4 + (\omega + \Delta)^2]} \qquad \text{Eq. 2}$$

$$\gamma_{1,\text{eff}} = \gamma_1 - \frac{4G^2|x_p|^2 \gamma_c \Delta \Omega_1}{[\gamma_c^2/4 + (\omega - \Delta)^2][\gamma_c^2/4 + (\omega + \Delta)^2]} \qquad \text{Eq. 3}$$

$$m\omega_1 \Omega_1 = m\omega_1^2 + 2S_1 - 24\frac{T_1 T_{1c}}{m\omega_1^2 + 4S_1}|x_p|^2 + 4F_{1c}|x_p|^2 \qquad \text{Eq. 4}$$

where $\Omega_1$ describes the frequency pulling of mode 1 due to $S_1$, $T_1$, and $F_{1c}$.

Mode coupling measurements for Device 1 are shown in Fig. 3a,b. Here mode 1 is probed while $\omega_p$ is swept from $\omega_2$ to $\omega_6$. The $|x_p|^2$ terms in equation (4) generate a downward frequency pulling of mode 1 when any mode is pumped directly on resonance; this is most visible at $\omega_p \approx 16$ MHz. Sideband cooling and amplification are also seen, and occur when pumping the red sideband of mode 6 and blue sideband of mode 2, respectively (Fig. 3a). Amplification also occurs at $\omega_p = 2\omega_1$, and is most notable at $V_{dc} = 10$ V, where $2\omega_1 \approx \omega_4$; this effect is studied in further detail in section S5 of the supplementary material.

The amplitude of mode 1 upon sideband cooling and amplification, shown in Fig. 3d, is nearly linear with pump amplitude – in contrast to the $|x_p|^2$ dependence predicted by equation (3). Analyzing the effective damping $\gamma_{1,\text{eff}}$ at the cavity sidebands reveals the source of this disagreement (Fig. 3e). Suppression of the sideband effects is observed around $\omega = \omega_1$, indicating a broadening of the sideband mode due to the probe amplitude $x_1$. For the case of $\omega_c = \omega_2$, $\omega_{sb} = \omega_6$, motion at $\omega_1$ and a non-zero coupling $T_{62}$ result in effective cooling of mode 6, hindering its ability to amplify mode 1. This quenching of the cavity effects can be avoided by probing mode 1 with lower amplitudes, and speaks to the dynamic range of a micromechanical filter/amplifier based on phonon cavity coupling; careful engineering of device modes such that $T_{sb,c} \approx 0$ would also counteract this effect. A detailed analysis of the measured damping $\omega_{1,\text{eff}}$ is presented in section S6 of the supplementary material.

Stronger phonon cavity effects have been measured in Device 2, where the larger device diameter permits the use of much weaker probe signals while maintaining comparable signal/noise. Measurements were performed with $V_{dc} = 4$V, so that $\omega_1 + \omega_2 \approx \omega_5$. Fig. 4a shows the membrane response upon pumping at $\omega_p = \omega_2 + \omega_1 = 2\pi \times 6.76$MHz with a voltage $v_p$ ramped from $0 - 400$mV$_{\text{pk}}$. Mode 1 is probed with $v = 0.4$mV$_{\text{pk}}$, and its motion undergoes amplification by a factor of 8.5 (19dB) before entering instability ($\gamma_{1,\text{eff}} \leq 0$) at $v_p = 300$mV$_{\text{pk}}$. Above this pump strength, mode 1 undergoes self-oscillation and locks onto the probe signal with a flat frequency response. The width of this flat region is 4kHz, significantly narrower than the unpumped linewidth, $\gamma_1/2\pi =$ 45kHz. Amplification of mode 1 continues to rise for higher pump strengths, ultimately reaching a factor of 18 (25dB).

In this configuration the graphene membrane also acts as a frequency mixer, generating motion at $\omega_p + \omega$ and $\omega_p - \omega$ (Fig. 4 b-c). Motion at $\omega_p - \omega \approx \omega_2$ signifies occupation of the cavity mode as a result of down-scattered pump phonons, and so is significantly larger (10 ×) than motion at $\omega_p + \omega$, where there is no mechanical resonance. Both of these mixed tones inherit the flat-top spectrum of mode 1 once it is in the self-oscillating regime.

Similar to driven motion in Device 1, red sideband pumping in Device 2 has been used to cool thermal motion of mode 1 to 200K (Fig. 4d). As in previous phonon cavity studies[4], the low cavity frequency $\omega_c \sim \omega_1$ (and high thermal phonon occupation) limits cooling in the all-mechanical system. Cooling motion towards the quantum ground state thus remains a task best suited for optical/microwave cavities, where $\omega_c \gg \omega_1$. However, interesting prospects arise if optical cavities and phonon cavities are utilized simultaneously to control graphene motion. For instance, optically cooling the phonon cavity enhances its capacity to mechanically cool the fundamental mode – in such a case cooling is limited only by the cooperativities $G^2|x_p|^2/\gamma_1\gamma_c$ of the two cavities. Moreover, the mechanical pump grants experimental control over the interaction strength of the two modes. Microwave-cavity-coupled graphene systems[16–18] are therefore ideal testbeds for quantum entanglement, squeezing, thermalization, and information exchange between modes near their ground state. The greatly enhanced $Q$ factors of graphene at dilution refrigerator temperatures[8,26] will only serve to strengthen these effects.

We have demonstrated tension-mediated coupling between mechanical modes in suspended graphene, and its potential for parametric control of this system. Sideband cooling and amplification of membrane motion, up to self-oscillation, have been observed within a single device. The potential for graphene membranes as frequency mixers with intrinsically flat pass-bands has also been shown. The coupling described is inherent in all graphene devices, and can be utilized to artificially enhance the $Q$'s of graphene-based sensors and electronics, and opens new possibilities in the study of coupled quantum systems.

**Methods**

Mechanical resonators were fabricated by growth of monolayer graphene through chemical vapor deposition and transfer to pre-patterned substrates. Prior to transfer, a supporting layer of 150nm Poly-methyl-methacrylate (PMMA) was spin-coated on the graphene surface and cured at 170°C. During transfer, the Cu growth substrate was etched using $FeCl_3$, and the PMMA/graphene film was cleaned by soaking in a series of deionized water baths After transfer, the film was coated with photo-resist and patterned via optical lithography; resist and PMMA were removed by submersion in N-methyl-2-pyrollidine at 80°C, releasing the suspended graphene membranes.

All measurements were performed at room temperature in a vacuum of $P < 10^{-6}$ mbar. Detection of mechanical motion was performed through optical interferometry, as detailed in previous work[11]. The light source used was a HeNe 633nm laser, focused to a spot of diameter ~ 1μm. Reflected light was monitored by a high-frequency photo-detector (New Focus 1811-FS) and recorded using a multi-channel lock-in amplifier (Zuirch Instruments HF2LI). The same lock-in amplifier was used to supply excitation voltages at the pump and probe frequencies. Graphene motion was inferred from the modulated laser power using the optical calibration scheme detailed in section 3 of the supplementary material.

**Acknowledgements**

The authors are grateful to P. Rose for assistance in growing the CVD graphene, and to D. MacNeill for insightful discussions and comments. This work was supported by the Cornell Center for Materials Research with funding from the NSF MRSEC program (grant no. DMR-1120296), and by Nanoelectronics Research Initiative (NRI) through the Institute for Nanoelectronics Discovery and Exploration (INDEX). Support was also provided by the Academy of Finland (through the project "Quantum properties of optomechanical systems"). Fabrication was performed in part at the Cornell NanoScale Facility, a member of the National Nanotechnology Coordinated Infrastructure, which is supported by the NSF (grant no. ECCS-15420819).

**Author contributions**

J.M.P., H.G.C., P.L.M., and R.D.A. designed the experiment; F.M. developed the supporting theory. R.D.A. and T.S.A. fabricated the samples; I.R.S. contributed to the design of the samples and the experimental set-up. R.D.A. and A.H. carried out the measurements. F.M. and R.D.A. analyzed the data. All authors discussed the results and commented on the manuscript.

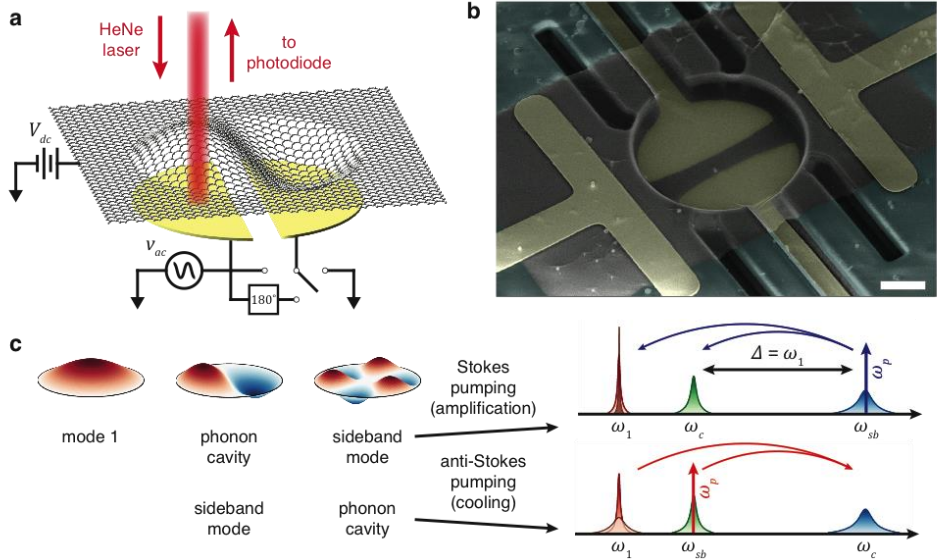

**Figure 1: The nonlinear system under test. a,** A cartoon of the experimental setup. Graphene motion is driven electrostatically by two metallic back-gates and detected through optical interferometry. The gates can be driven in various configurations to favor excitation of the fundamental mode, higher frequency modes, or both. **b,** False-color electron micrograph of Device 1; scale bar is 2µm. **c,** Schematic of the three modes necessary for efficient sideband pumping and their relative positions in frequency space. Curved arrows indicate the direction of energy flow when the system is pumped at $\omega_p$.

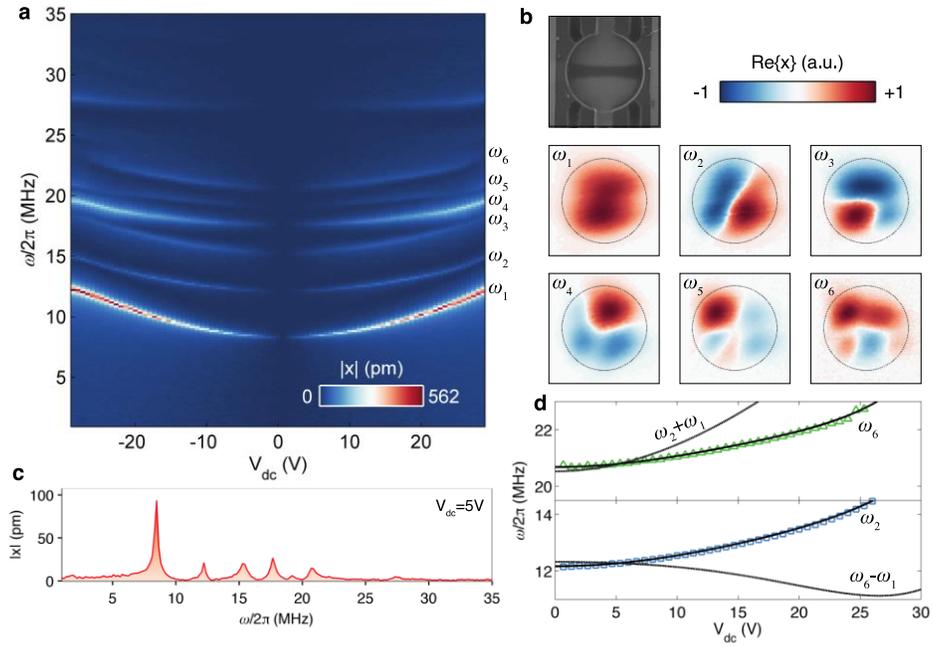

**Figure 2: Multimode membrane characterization. a,** Frequency dispersion with $V_{dc}$ for lowest 6 modes in Device 1. **b,** Mechanical mode shapes at $V_{dc} = 5V$ measured by scanning the detection laser across the membrane surface. The electron micrograph is given as a reference for orientation. **c,** Frequency spectrum at $V_{dc} = 5V$. **d,** Resonant frequencies of mode 2 and mode 6 extracted from **a**, in comparison to their sidebands with mode 1. Appreciable overlap between these frequencies occurs for $V_{dc} = 0 - 7.5V$, and so strong phonon cavity effects are expected in this range.

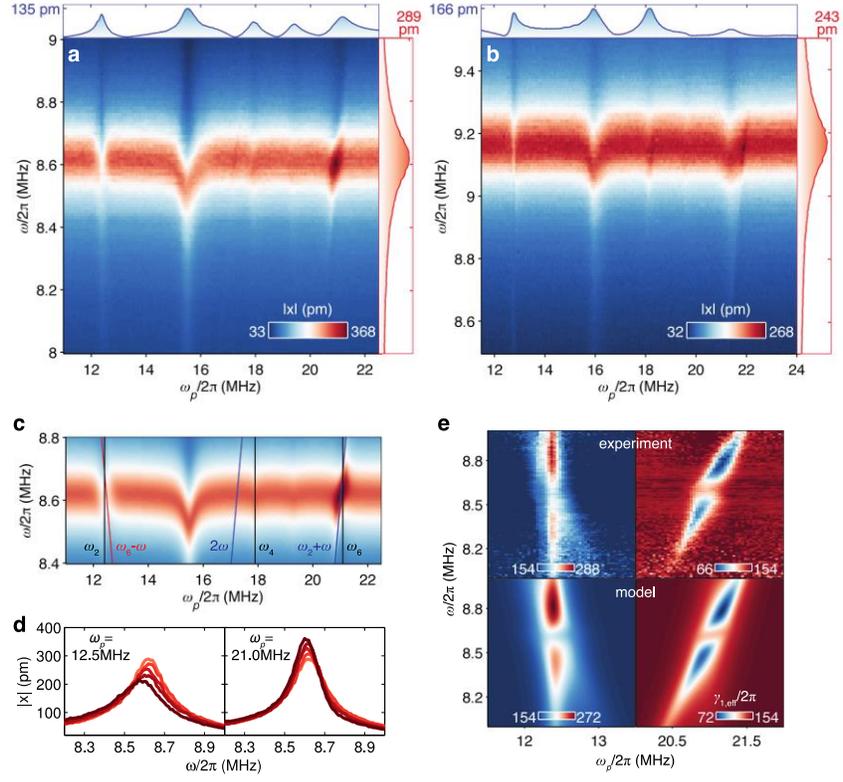

**Figure 3: Phonon pumping in Device 1. a,b,** Main panels: mode 1 amplitude vs pump frequency $\omega_p$ and probe frequency $\omega$ at $V_{dc} = 5V$ and 10V, respectively. Right panels: vertical slices through the data at the highest $\omega_p$ value. Upper panels: motion in the membrane at $\omega_p$, measured simultaneously with the main panel. Measurements for both $V_{dc}$ values were performed with equal excitation forces ($F \propto V_{dc} v_{ac}$) at the pump frequency (and at the probe frequency). Cavity cooling and amplification of mode 1 are stronger in **a**, where there is better mode-sideband alignment. **c,** Modeled behavior in **a** based on Equations (2-4). Solid lines denote relevant frequencies for sideband cooling and amplification. **d,** Measured cooling and amplification at $V_{dc} = 5V$ for linearly increasing pump strength (darkening lines). **e,** Effective mode 1 damping as measured in **a** (top) and modeled by equation (3) (bottom) in kHz; colors in the upper panel are truncated to the intrinsic damping $\gamma_1/2\pi = 154$ kHz. Quenching of the cavity effect near $\omega = \omega_1$ is due to the large mode 1 amplitude (and nonzero $T_{sb,c}$). Only two free parameters ($T_{1c}$ and $T_{sb,c}$) were used to produce each of the lower panels.

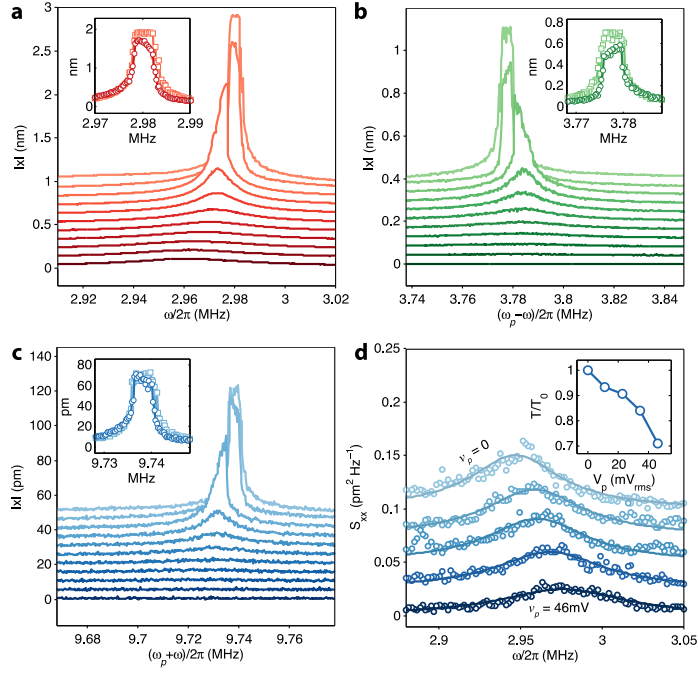

**Figure 4: Parametric self-oscillation and cooling in Device 2. a,** Amplification of mode 1 ($\omega_1/2\pi = 3.0$MHz, $\gamma_1/2\pi = 45$kHz) and transition to mechanical lasing ($\gamma_{1,\text{eff}} \leq 0$) via mode coupling. Mode 1 is probed with a weak drive ($v = 0.4$mV$_{\text{pk}}$) as mode 2 is pumped at its Stokes sideband ($\omega_p = 6.8$MHz) with increasing pump strength ($v_p = 0 - 400$mV$_{\text{pk}}$). Curves are vertically offset for clarity. Inset: Saturation of vibrational amplitude and "flat-top" response of the self-oscillating mode; no vertical offset is applied. **b,c,** Frequency mixing via mechanics. Measured membrane motion at $\omega_p - \omega$ and $\omega_p + \omega$, recorded simultaneously with **a**. **d,** Measured spectral noise density near $\omega_1$ upon pumping the anti-Stokes sideband of mode 5 ($\omega_p = 3.8$MHz). Curves are vertically offset for clarity. Inset: Effective temperature of mode 1, corresponding to the area under the $S_{xx}$ fits. The frequency spectrum of Device 2 is given in section S2 of the supplementary material.

# Tunable phonon cavity coupling in graphene membranes
## Supplementary Information

R. De Alba[1], F. Massel[2], I. R. Storch[1], T. S. Abhilash[1], A. Hui[3], P. L. McEuen[1,4], H. G. Craighead[3] & J. M. Parpia[1]

[1]Department of Physics, Cornell University, Ithaca, New York 14853, USA
[2]Department of Physics, Nanoscience Center, University of Jyväskylä, F1-40014, Finland
[3]School of Applied & Engineering Physics, Cornell University, Ithaca, New York 14853, USA
[4]Kavli Institute at Cornell for Nanoscale Science, Ithaca, New York 14853, USA

**Outline:**

## S1     Mode coupling in a 2D circular membrane with electrostatic drive – Theory

The static deformation and the dynamics of the circular membrane are described in terms of the following Lagrangian:

$$L = \frac{\rho}{2} \int dA\, \dot{x}^2 - \frac{D}{2} \int dA\, (\Delta x)^2 - \frac{C}{16} \left[ \frac{1}{A} \int dA\, (\nabla x)^2 \right] \cdot \int dA\, (\nabla x)^2 \\ - \frac{T_0}{2} \int dA\, (\nabla x)^2 - \frac{\epsilon_0 V_g^2}{2} \int \frac{dA}{d-x} \tag{S1}$$

where $\rho$ is the surface density, $D$ is the flexural rigidity, $C$ the in-plane stiffness, $T_0$ the built-in tension, $V_g$ the gate voltage, $d$ the gate-sheet separation, $x$ the deformation. If the description of the system can be performed in terms of a continuum model, we have

$$C = \frac{Eh}{1-\nu^2} \tag{S2}$$

$$D = \frac{Eh^3}{12(1-\nu^2)} \tag{S3}$$

The first term in Equation S1 represents the kinetic energy associated with the dynamics of the membrane, the second its flexural elastic energy, the third the energy associated with the deformation-induced tension (treated on a mean-field level), the fourth the energy due to built-in tension and last term corresponds to the capacitive coupling with the back gate.

By expanding $x$ into static and resonant components $x(\vec{r}, t) = x_0(\vec{r}) + \sum_i x_i(t) \xi_i(\vec{r})$ – where $\xi(\vec{r})$ is the dimensionless, normalized profile of mode $i$ – it is straightforward to show that Equation S1 leads to the Hamiltonian given in Equation 1 of the main text, with nonlinear coefficients

$$L_i = \left[ \frac{T_0}{2} + \frac{C}{4A} \int dA(\nabla x_0)^2 \right] \cdot \int dA(\nabla x_0 \nabla \xi_i) \tag{S4}$$

$$S_i = \frac{C}{4A} \left[ \int dA(\nabla x_0 \nabla \xi_i) \right]^2 \tag{S5}$$

$$T_i = \frac{C}{4A} \left[ \int dA(\nabla x_0 \nabla \xi_i) \right] \cdot \int dA(\nabla \xi_i)^2 \tag{S6}$$

$$F_i = \frac{C}{8A} \left[ \int dA(\nabla \xi_i)^2 \right]^2 \tag{S7}$$

$$S_{ij} = \frac{C}{8A} \left[ \int dA(\nabla x_0 \nabla \xi_i) \right] \cdot \int dA(\nabla x_0 \nabla \xi_j) \tag{S8}$$

$$T_{ij} = \frac{C}{4A} \left[ \int dA(\nabla x_0 \nabla \xi_i) \right] \cdot \int dA(\nabla \xi_j)^2 \tag{S9}$$

$$F_{ij} = \frac{C}{16A} \left[ \int dA(\nabla \xi_i)^2 \right] \cdot \int dA(\nabla \xi_j)^2. \tag{S10}$$

Above we have assumed displacements are large such that the flexural rigidity is negligible: $x \gg h$. We now proceed to calculate these coefficients explicitly for a circular membrane geometry.

It is important to note that the terms listed in Equations S4-S10 and included in Equation 1 of the main text are the only nonlinear terms expected from the perspective of the mean field

From Equation S1 it is possible to derive an equation for the static displacement of the membrane

$$\Delta^2 \zeta_0 - \tau \Delta \zeta_0 = \kappa_0 V_g^2 \tag{S11}$$

where $\zeta_0$ represents the static displacement (expressed in units of $r_0$) of the membrane in presence of a time-independent external voltage $V_g$, $\kappa_0 = \epsilon_0 r_0^3/[2(d-x)^2 D]$ and $\tau\ (= T r_0^2/D)$ is the (dimensionless) membrane tension, whose value has to be determined from the solution of the following equation

$$\tau = \tau_0 + \frac{2\epsilon}{\pi} \int d\Omega\ (\nabla \zeta_0)^2. \tag{S12}$$

The solution of Equation S11 is given by

$$\zeta_0 = \frac{\kappa_0 V_g^2}{4\tau} \left[ 1 - r^2 + 2 \frac{I_0(\sqrt{\tau} r) - I_0(\sqrt{\tau})}{\tau I_1(\sqrt{\tau})} \right]. \tag{S13}$$

In order to determine $\tau$, the result given in Equation S13 is substituted into Equation S12, yielding the following self-consistent equation for $\tau$

$$\tau = \tau_0 + \frac{\epsilon \kappa_0^2 V_g^4}{4\tau^2} \left( \frac{16}{\tau} - \frac{2 R_0(\tau)}{\sqrt{\tau}} - R_0^2(\tau) + 3 \right) \tag{S14}$$

with $R_0(\tau) = I_0(\sqrt{\tau})/I_1(\sqrt{\tau})$ ($I_n(x)$ is the modified Bessel function of the first kind of order $n$). From the adimensionalized version of the Lagrangian given in Equation S1, it is possible to obtain the Hamiltonian $\mathcal{H}$ describing the dynamics of small oscillations around $\zeta_0$ in terms of the operators $a_i$, $a_i^\dagger$, $X \doteq a_i + a_i^\dagger$

$$\begin{aligned}
\mathcal{H} = &\ \tilde{\omega}_a \hat{a}_a^\dagger \hat{a}_a + \tilde{\omega}_b \hat{a}_b^\dagger \hat{a}_b \\
&+ \mathcal{S}_a \hat{X}_a^2 + \mathcal{L}_a \hat{X}_a + \mathcal{T}_a \hat{X}_a^3 + \mathcal{F}_a \hat{X}_a^4 \\
&+ \mathcal{S}_b \hat{X}_b^2 + \mathcal{L}_b \hat{X}_b + \mathcal{T}_b \hat{X}_b^3 + \mathcal{F}_b \hat{X}_b^4 \\
&+ \mathcal{T}_{ab} \hat{X}_b^2 \hat{X}_a + \mathcal{T}_{ba} \hat{X}_a^2 \hat{X}_b + \mathcal{F}_{ab} \hat{X}_a^2 \hat{X}_b^2
\end{aligned} \tag{S15}$$

Where

$$\mathcal{L}_i = \frac{T}{D} \bar{B}_i(T) \tilde{x}_i \tag{S16}$$

$$\mathcal{S}_i = \frac{\epsilon}{\pi} \bar{B}_i^2(T) \tilde{x}_i^2 \tag{S17}$$

$$\mathcal{T}_i = \frac{2\epsilon}{\pi} \alpha_i^2 \bar{B}_i(T) \tilde{x}_i^3 \tag{S18}$$

$$\mathcal{F}_i = \frac{\epsilon}{2\pi} \alpha_i^4 \tilde{x}_i^4 \tag{S19}$$

$$\mathcal{T}_{ij} = \frac{2\epsilon}{\pi} \alpha_j^2 \bar{B}_i(T) \tilde{x}_j^2 \tilde{x}_i \tag{S20}$$

$$\mathcal{F}_{ij} = \frac{\epsilon}{2\pi} \alpha_i^2 \alpha_j^2 \tilde{x}_i^2 \tilde{x}_j^2 \tag{S21}$$

with $i, j \in \{a, b\}$, and $\tilde{x}_i = \sqrt{\frac{1}{2\mu \tilde{\omega}_i}}$. Moreover we have

$$\bar{B}_i(T) = B_i(\tau) = \frac{\sqrt{\pi} \kappa_0 V_g^2}{\tau} \alpha_i \left[ \frac{2}{\alpha_i^2} - \frac{\sqrt{\tau}}{\tau + \alpha_i^2} R_0(\tau) \right] \tag{S22}$$

and

$$\epsilon = \frac{3r_0^2}{2h^2} \tag{S23}$$

$$\tilde{\omega}_i = \frac{\alpha_i \hbar}{r_0 D} \sqrt{\frac{T}{\rho}} \tag{S24}$$

$$\mu = \frac{\pi \rho r_0^4 D}{\hbar^2}. \tag{S25}$$

The dimensionless coefficients $\mathcal{L}_i, \mathcal{S}_i, \mathcal{T}_i, \mathcal{F}_i$ can be re-dimensionalized according to: $L_i = \mathcal{L}_i D/(r_0 \tilde{x}_i)$, $S_i = \mathcal{S}_i D/(r_0 \tilde{x}_i)^2$, $T_{ij} = \mathcal{T}_i D/(r_0^3 \tilde{x}_j^2 \tilde{x}_i)$, and $F_{ij} = \mathcal{F}_i D/(r_0^4 \tilde{x}_i^2 \tilde{x}_j^2)$. Moreover, in Equation S15 and the discussion that follows, the second order coupling $S_{ij}$ has been excluded as only the fundamental mode has appreciable overlap with the static deformation $x_0$.

For large values of the induced tension the values of $T$ and $\bar{B}_i(T)$ are given by

$$T = \frac{1}{4} \left[ \frac{r_0^2 \epsilon_0^2 C}{(d-x)^4} \right]^{1/3} V_g^{4/3} \tag{S26}$$

$$\bar{B}_i(T) = \frac{4\sqrt{\pi}}{\alpha_i} \left[ \frac{\epsilon_0 r_0}{(d-x)^2 C} \right]^{1/3} V_g^{2/3}. \tag{S27}$$

In Equation S15 $L_i$ is a term that can be trivially "displaced" away, $T_{ij}, T_i$ and $F_i, F_{ij}$ are the term relevant for the radiation-pressure and Duffing physics, while $S_i$ represents a shift in frequency of the mode considered.

As an example, we focus our attention on 3 different modes of a circular membrane: $(0,1)$ (hereafter mode a), $(1,1)$ (mode b for the red-detuned case), $(3,1)$ (mode b for the blue-detuned case). This choice is related to the necessity of having three modes for which $\omega_1 \simeq \omega_2 + \omega_3$. This condition is optimal in terms of radiation pressure-like coupling between modes. The $(0,1)$ mode plays the role of the mechanical mode, while modes $(1,1), (3,1)$ play the role of the driving tone (cavity) and cavity (driving tone) for red- (blue-) sideband detuning respectively. Due to the large density of states at the cavity resonance, driving the system close to one of its resonances (which has non-negligible overlap with the cavity resonance) allows for an efficient excitation of the sideband, leading to a stronger optomechanical coupling, for a given input drive, as compared to the case for which the resonance close to the pump frequency is absent.

The physics leading to the frequency and damping shift of the fundamental mode, can be essentially explained with the same analysis performed for optomechanical systems. The driving around $\omega_p$ can be interpreted as the "optical pump", detuned away from a "cavity" by the mechanical resonant frequency.

The calculation goes as follows: the strong drive around $\omega_p$ field $\beta_{in}$ is determined by the solution of the I/O equations for a free (i.e. uncoupled to other modes) mode, this mode will then be considered as the sideband (with respect to $\omega_c$) drive in the I/O equations for the coupled system whose unitary dynamics is described by the Hamiltonian Equation S15. The analysis of these quantum Langevin equations (QLEs), will be performed in terms of a standard linearisation procedure, in complete analogy to what is done in the context of optomechanical systems.

With the approximations mentioned above, the cavity field around $\omega_p$ can be written as

$$\beta = \frac{\sqrt{\gamma_p}\beta_{in}}{\frac{\gamma_p}{2} - i(\omega - \omega_p)} \quad (S28)$$

The value of $\beta$ represents thus, on one hand, the oscillation amplitude when the resonator is driven close to the resonance $\omega_p$, and, on the other the amplitude of oscillations at a frequency which is detuned by $\omega_p - \omega_c \simeq \omega_f$ ($\omega_f = \omega_a$). The relative values of $\omega_a$, $\omega_{b1}$ and $\omega_{b2}$ allow us therefore to have a strong field $\beta$ since we are driving the system on resonance, and at the same time, exploit the optomechanical-like sideband physics.

In order to describe the nonlinear sideband physics, we write the QLEs associated with the Hamiltonian Equation S15

$$\begin{aligned}\dot{a} &= -i\omega_a a - i\mathcal{L}_a - i2\mathcal{S}_a(a^\dagger + a) - i3\mathcal{T}_a(a^\dagger + a)^2 - i4\mathcal{F}_a(a^\dagger + a)^3 \\ &\quad -i\mathcal{T}_{ab}(b^\dagger + b)^2 - i2\mathcal{F}_{ab}(a^\dagger + a)(b^\dagger + b)^2 - \frac{\gamma_a}{2}a + \sqrt{\gamma_a}a_{in}\end{aligned} \quad (S29)$$

$$\begin{aligned}\dot{b} &= -i\omega_b b - i4\mathcal{F}_b(b^\dagger + b)^3 \\ &\quad -i\mathcal{T}_{ab}(a^\dagger + a)(b^\dagger + b) - i2\mathcal{F}_{ab}(a^\dagger + a)^2(b^\dagger + b) - \frac{\gamma_b}{2}b + \sqrt{\gamma_b}b_{in}.\end{aligned} \quad (S30)$$

We can solve Equations S29,S30 perturbatively, assuming that we can expand $a$ and $b$ as $a \to \alpha + a$ and $b \to \beta + b$, where $\alpha$ represents the coherent oscillation amplitude of the fundamental mode induced by $\beta$ whose value is given by Equation S28. The value of $\alpha$ can be obtained as the solution of the zeroth-order term in the expansion of Equation S29, which can be written as

$$\begin{aligned}\dot{\alpha} &= -i\omega_a \alpha - i\mathcal{L}_a - i2\mathcal{S}_a(\alpha^* + \alpha) - i3\mathcal{T}_a(\alpha^* + \alpha)^2 - i4\mathcal{F}_a(\alpha^* + \alpha)^3 \\ &\quad -i\mathcal{T}_{ab}(\beta^* + \beta)^2 - i2\mathcal{F}_{ab}(\alpha^* + \alpha)(\beta^* + \beta)^2 - \frac{\gamma_a}{2}\alpha + \sqrt{\gamma_a}\alpha_{in}.\end{aligned} \quad (S31)$$

In the substitution $a \to \alpha + a$ we have assumed that $\dot{\alpha} = 0$. This assumption is justified when the conditions $\omega_a < 2\omega_b$, and $\gamma_a < (\omega_a - \omega_b)$ are fulfilled, (rotating wave approximation), leading to

$$\alpha = -\frac{2\mathcal{T}_{ab}}{\omega_a + 4\mathcal{S}_a}|\beta|^2 \quad (S32)$$

where higher-order terms have been neglected, and we have assumed, without loss of generality $\alpha^* = \alpha$.

The first-order term in the expansion of Equations S29,S30 can be written as

$$\begin{aligned}\dot{a} &= -i\omega_a a - i\mathcal{L}_a - i\mathcal{S}_a(a^\dagger + a) - i6\mathcal{T}_a(\alpha^* + \alpha)(a^\dagger + a) - i12\mathcal{F}_a(\alpha^* + \alpha)^2(a^\dagger + a) \\ &\quad -i2\mathcal{T}_{ab}(\beta^* b + \beta b^\dagger) - i2\mathcal{F}_{ab}[(\beta^* + \beta)^2(a^\dagger + a) + (\alpha^* + \alpha)(\beta^* + \beta)(b^\dagger + b)] \\ &\quad -\frac{\gamma_a}{2}a + \sqrt{\gamma_a}a_{in}\end{aligned} \quad (S33)$$

$$\begin{aligned}\dot{b} &= -i\omega_b b - i3\mathcal{F}_b(\beta^* + \beta)(b^\dagger + b) \\ &\quad -i\mathcal{T}_{ab}[(\beta^* + \beta)^2(a^\dagger + a) + (\alpha^* + \alpha)(\beta^* + \beta)(a^\dagger + a)] \\ &\quad -i\mathcal{F}_{ab}[2(\alpha^* + \alpha)^2(b^\dagger + b) + (\alpha^* + \alpha)(\beta^* + \beta)(a^\dagger + a)] - \frac{\gamma_b}{2}b + \sqrt{\gamma_b}b_{in}.\end{aligned} \quad (S34)$$

Neglecting again higher-order terms, Equations S33,S34 can be written as

$$\begin{aligned}\dot{a} &= -i\omega_a a - i\left[\mathcal{S}_a - 24\frac{\mathcal{T}_a\mathcal{T}_{ab}}{\omega_a + 2\mathcal{S}_a}|\beta|^2 + 4\mathcal{F}_{ab}|\beta|^2\right]a \\ &\quad -i2\mathcal{T}_{ab}(\beta^* b + \beta b^\dagger) - \frac{\gamma_a}{2}a + \sqrt{\gamma_a}a_{in}\end{aligned} \quad (S35)$$

$$\dot{b} = -i\omega_b b - i2\mathcal{T}_{ab}\beta(a^\dagger + a) - \frac{\gamma_b}{2}b + \sqrt{\gamma_b}b_{in} \tag{S36}$$

where RWA has been used for $a$, $b$ and $\beta$. Equations S35,S36 can be written more compactly as

$$\dot{a} = -i\Omega_a a - i(G^*b + Gb^\dagger) - \frac{\gamma_a}{2}a + \sqrt{\gamma_a}a_{in} \tag{S37}$$

$$\dot{b} = -i\omega_b b - iG(a^\dagger + a) - \frac{\gamma_b}{2}b + \sqrt{\gamma_b}b_{in} \tag{S38}$$

Where

$$\Omega_a = \omega_a + 2\mathcal{S}_a - 24\frac{\mathcal{T}_a\mathcal{T}_{ab}}{\omega_a + 2\mathcal{S}_a}|\beta|^2 + 4\mathcal{F}_{ab}|\beta|^2 \tag{S39}$$

$$G = 2\mathcal{T}_{ab}\beta. \tag{S40}$$

Equations S37,S38 are the equation of motion of two linearly coupled harmonic oscillators, and are equivalent to the linearised equation of motion for an optomechanical system. It can be shown that in this setup the mode a undergoes a frequency shift and a damping shift given by

$$\omega_{\text{eff}} = \sqrt{\Omega_a^2 + \frac{|G|^2 \Delta\Omega_a [\gamma_b^2/4 - \omega^2 + \Delta^2]}{2[\gamma_b^2/4 + (\omega - \Delta)^2][\gamma_b^2/4 + (\omega + \Delta)^2]}} \tag{S41}$$

$$\gamma_{\text{eff}} = \gamma_a - \frac{\gamma_a |G|^2 \Delta\Omega_a}{[\gamma_b^2/4 + (\omega - \Delta)^2][\gamma_b^2/4 + (\omega + \Delta)^2]}. \tag{S42}$$

From Equations S37 and S41, it is clear how the observed frequency shifts have 2 different sources. On the one hand, it is determined by "geometric" nonlinearities, i.e. effects which are essentially determined by the eigenmode shapes, dictating the value of $\Omega_a$ in Equations S39, on the other it depends on the mechanical analogue of optomechanical effects $\omega_{\text{eff}}$.

Even though residual nonuniform tensions in the device (as evidenced by the mode shapes presented in Figure S2b of the main text) do not allow for a direct quantitative comparison, the description adopted here for the exact solution of the circular membrane can be applied to the experimental results provided the coefficients given in Equation S15 are correctly identified from the experiment.

## S2 Characterization of Devices 1 and 2

The graphene devices studied in this work are shown in Figure S1. Their physical properties are given in Table S1. The mass density of these membranes is ~ 10× that of bare graphene due to surface contaminants (most likely PMMA from fabrication[1]). Mass density here has been measured by fits to the AC amplitude of motion as $V_{dc}$ is varied, described in Section S3 and presented in Figure S3. These values can also be obtained from fits to the resonant frequency dispersion $f(V_{dc})$, as has been described in numerous works previously[2–5]. The intrinsic tension $T_0$ is calculated from $f = (\alpha/2\pi r_0)\sqrt{T/\rho}$, where $\alpha \approx 2.404$.

A spectrum for Device 2 at $V_{dc} = 4V$ is given in Figure S2. The pumping conditions used in Figure 4 of the main text are also shown.

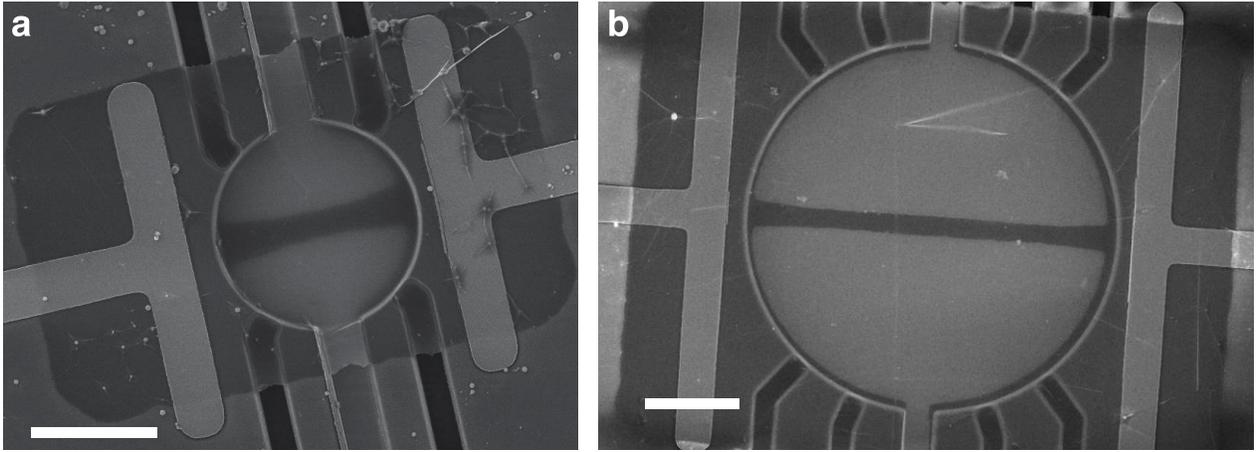

**Figure S1: Devices studied. a,b,** Scanning electron micrograph of Device 1 and 2, respectively. Scale bars are 5μm. In both cases graphene is suspended above a 1.7μm-deep circular trench in $SiO_2$. Linear trenches (6 in **a** and 10 in **b**) allow fluid to drain from under the graphene during device fabrication. All but two trenches terminate in a thin $SiO_2$ bridge so as not to affect the membrane boundary conditions. The remaining two trenches carry 50nm-thick platinum leads to the split back-gates. Platinum source and drain leads contact the graphene bottom surface.

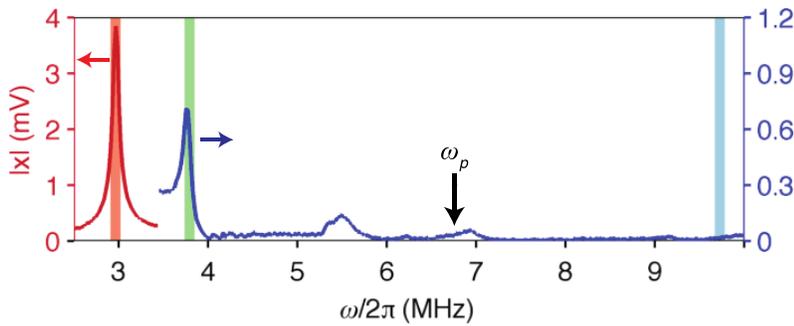

**Figure S2: Spectrum of Device 2.** The pump configuration used to obtain Figure 4a-c is shown. Vertical bands denote the three frequency ranges in which motion was measured while pumping $\omega_p$.

**Table S1: Graphene device properties**

| Device # | diameter (μm) | $\rho/\rho_{graphene}$ | $f_1(V_{dc}=0)$ (MHz) | $T_0$ (N/m) |
|---|---|---|---|---|
| 1 | 7.8 | 11 ± 2 | 8.35 | 0.060 |
| 2 | 19.9 | 9.5 ± 1 | 2.9 | 0.040 |

# S3 Calibration of optical detection system

Calibration of the electrical and optical components of our setup were performed by pulling on Device 1 with a varying DC gate voltage (and fixed AC voltage) while measuring both the AC and DC components of our reflected laser power. As described below, this process allows us to determine the absolute deflection of our graphene membrane (both the static and resonant components), as well as the effective AC gate voltage that is "felt" by the graphene, $v_{ac}$. This latter value is substantially smaller than the applied AC voltage, $V_{ac}$, due to parasitic losses of our cables and wire bonds, contact resistance of the graphene, and other unavoidable losses. The DC gate voltage, $V_{dc}$, does not suffer this effect, as the graphene-gate capacitor ($C$) will reach the experimentally applied voltage within a few $RC$ time constants, where $R$ encompasses all series resistances.

Following an approach reported previously[4], the graphene is considered to be situated in an optical standing wave generated by the incident laser light and reflection from the metallic back-gate. Because of the graphene's 2.3% optical absorption[6], the overall reflected power out of the system is sensitive to the graphene position within this standing wave; this sensitivity enables us to detect graphene motion. The DC component of our reflected laser power depends on graphene position, $x$, as

$$P_{dc} = P_0 + \Delta P \sin\left(\frac{4\pi}{\lambda}x + \theta\right) \qquad (S43)$$

where $\lambda$ is the wavelength of light used (633nm) and $P_0$, $\Delta P$, $\theta$ are the average power, modulation depth, and phase of the standing wave at the graphene, respectively. The position $x$ can be altered by pulling the graphene towards the back-gate with a bias voltage, $V = V_{dc} + v_{ac} \sin \omega t$. If $v_{ac} \ll V_{dc}$, and $\omega$ is far below any mechanical resonance of the graphene, the ac portion of our reflected laser power is:

$$P_{ac} = \Delta P \cos\left(\frac{4\pi}{\lambda}x + \theta\right) \cdot \frac{4\pi}{\lambda} \frac{dx}{dV_{dc}} v_{ac}. \qquad (S44)$$

In principle $P_0$, $\Delta P$, and $\theta$ can be calculated from the incident power, refractive index of graphene and back-gate, and the graphene-gate separation. However, this calculation is complicated by: (1) the quality and size of our laser spot (diameter $\approx$ 1μm), (2) the thickness, roughness, and refractive index of surface contaminants on the graphene[1], and (3) graphene adherence to the vertical walls of the trench[7], among other uncertainties. These parameters should therefore be measured experimentally – in this case by pulling the graphene a significant fraction of the distance $\lambda/4$. To simplify calculations, the membrane deflection profile is assumed to be a paraboloid $x(r) = x_0(1 - r^2/r_0^2)$, where $r$, $r_0$ are radial position and membrane radius, and $x_0$ is the height of the membrane center. The membrane position can then be calculated from the balance of tension and electrostatic forces:

$$4\pi T_0 x_0 + \frac{4\pi E}{r_0^2} x_0^3 = \frac{1}{2} C' V_{dc}^2. \qquad (S45)$$

Above, $T_0$ and $E$ are the 2D membrane tension and 2D Young's modulus (in N/m), respectively, and $C' = dC/dx$ where $C$ is the graphene-gate capacitance. This is a modified version of force equations reported elsewhere[4,7], adjusted for the paraboloid approximation. The tension $T_0$ can be recast as the membrane mass density $\rho$ by knowledge of the resonant frequency: $\rho = T_0(2.404/\omega_1 r_0)^2$.

Equations S43-S45 thus provide a means to model our optical system as the bias voltage $V_{dc}$ is varied. Figure S3 shows a representative data set from which $P_0$, $\Delta P$, $\theta$, $\rho$, and $v_{ac}/V_{ac}$ are measured for Device 1. Here, $V_{dc}$ is swept ($0 - 35$V) while a constant $V_{ac}$ (200mV$_{pk}$, $\omega = 2\pi \times 100$kHz) is applied to both back-gates. AC data is fitted first, by numerically finding the roots of Equation S45 and applying them to Equation S44; results are shown in Figure S3a. This fit provides values for $\rho$, $\Delta P \cdot v_{ac}$, and $\theta$. With these parameters determined, $P_{dc}$ is fitted to Equation S44 to obtain $P_0$ and $\Delta P$; this is shown in Figure

S3b. The excellent agreement of this second, more constrained fit verifies the validity of this model. With all the parameters of Equations S43-S45 determined, we can plot the DC and AC membrane deflection for this data set, as shown in Figure S3c-d. Interesting features in these curves are: 1) In Figure S3d, the transition from quadratic to sub-linear DC deflection above $V_{dc} \approx 25\text{V}$ caused by the $E$ term in Equation S45. 2) The resulting maximum in AC deflection that this transition produces, as shown in Figure S3c.

In performing these fits, the measured modulus of similarly produced CVD graphene[8] is used, $E = 55\text{N/m}$. Moreover, for the purposes of Equations S43 and S44, it is assumed that our laser spot performs a "point-like" measurement of $x$ at the membrane's center of mass, $r_{cm} = r_0/\sqrt{2}$. Equation S45 was corrected for this off-center measurement, $x(r_{cm}) = x_0/2$. The resulting mass density of Device 1 is $\rho/\rho_g = 11 \pm 2$, where $\rho_g = 0.75\text{mg/m}^2$ is the density of monolayer graphene; the extra mass is attributed to polymer contaminants from fabrication. The AC gate voltage "felt" by the graphene is $v_{ac} = 11\text{mV}_{\text{pk}}$, or $v_{ac}/V_{ac} = 5.5\%$. From similar fits, Device 2 is found to have $\rho/\rho_g = 9.5$ and $v_{ac}/V_{ac} = 3.8\%$.

In the main text, resonant motion is converted from μV (generated by our photodiode) to pm using the above measured values of $\rho$ and $v_{ac}/V_{ac}$ to calculate the applied force during any measurement, and the resulting motion of the fundamental mode. This value is then compared to the measured amplitude (in μV) on resonance. It should be noted that because this calibration uses only mode 1 as a reference, the relative amplitude measured for each of the higher modes depends upon the position of our laser spot.

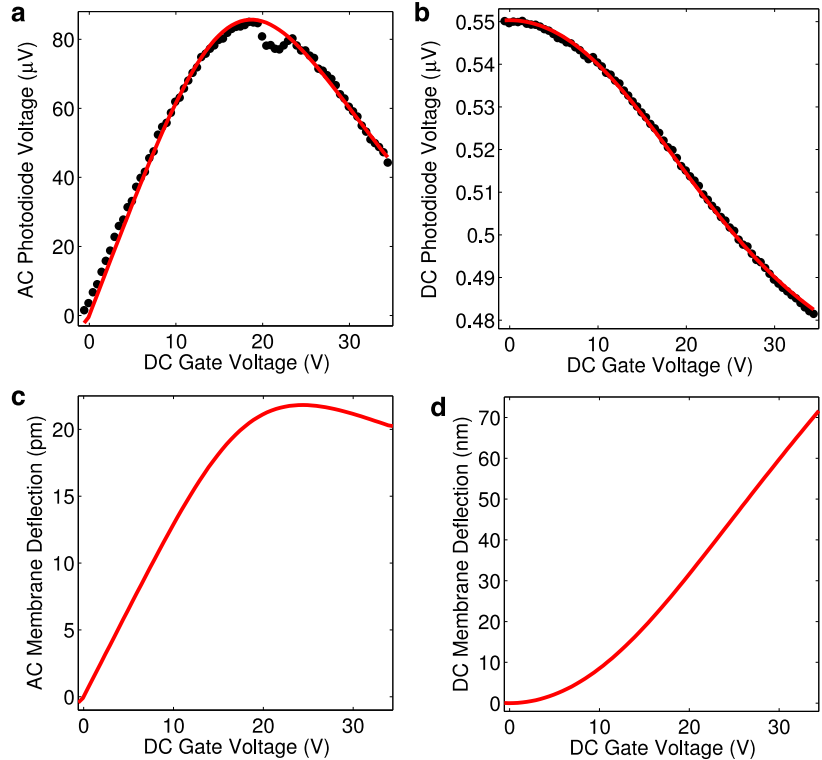

**Figure S3: a,** Measured ac reflected laser power as graphene is driven far below resonance at fixed $V_{ac}$ and varying $V_{dc}$. Red line: Three-parameter fit, as described in the text. **b,** Measured dc reflected laser power collected in synchrony with **a** (black points). Red line: two-parameter fit, using values taken from fit to **a**. **c,d,** Calculated membrane deflection at $\omega = 2\pi \times 100\text{kHz}$ and $V_{ac} = 200\text{mV}_{\text{pk}}$ resulting from the fits in **a** and **b**. The responsivity and transimpedance gain specified for our photodiode (New Focus 1801-fs-ac) are used to convert between measured voltage and input laser power.

## S4   Measurement of effective damping and frequency

In order to fully characterize the effects of mode coupling on the frequency and damping of mode 1, its response amplitude $|x_1|$ and phase $\phi_1$ must both be measured. Below we show that when compared to the amplitude and phase of a "reference state" of the same resonator, these two numbers can be used to infer the effective damping and resonant frequency of the mode.

The equation of motion for mode 1 in the absence of nonlinearities is

$$\ddot{x} + \gamma\dot{x} + \omega_0^2 x = \frac{F}{m}e^{i\omega t} \tag{S46}$$

where the '1' subscripts have been excluded for brevity and $\omega_0, \gamma$ are the natural frequency and damping. $F, m$ are the applied electrostatic force and membrane mass. In the presence of mode coupling and Stokes (or anti-Stokes) pumping, only $\omega_0$ and $\gamma$ are altered:

$$\ddot{x} + \gamma_{\text{eff}}\dot{x} + \omega_{0,\text{eff}}^2 x = \frac{F}{m}e^{i\omega t}. \tag{S47}$$

Assuming $F$ is real, the driven response of mode 1 then becomes

$$x(t) = \frac{F/m}{\omega_{0,\text{eff}}^2 - \omega^2 + i\gamma_{\text{eff}}\omega} e^{i\omega t}. \tag{S48}$$

Expressing $x(t)$ by its quadratures $\text{Re}\{x\} = X \cos \omega t$ and $\text{Im}\{x\} = Y \sin \omega t$, and amplitude $|x| = R$, the effective resonant frequency and damping can be determined from:

$$\omega_{0,\text{eff}}^2 = \omega^2 + \frac{FX}{mR^2} \tag{S49}$$

$$\gamma_{\text{eff}} = -\frac{FY}{m\omega R^2}. \tag{S50}$$

In generating Figure 3e of the main text, motion was calibrated based on a "reference" region of $(\omega, \omega_p)$ space where mode coupling effects are negligible: $\omega_p/2\pi = 22 - 22.5$ MHz. Data from this region was fitted to Equation S48 to calibrate the amplitude of motion $R$ (according to Section S2), as well as adjust the measured phase such that $X = 0$ on resonance. Equations S49-S50 were then used to convert $X, Y$ to $\omega_{1,\text{eff}}, \gamma_{1,\text{eff}}$ for each point in $(\omega, \omega_p)$ space, as shown in Figure S4.

It should be noted that Equations S49-S50 can easily be modified to account for a Duffing nonlinearity in the "reference" region. Slight variations in the optical detection efficiency can also be modeled quite effectively, as was necessary for Figures S4c,d & 3e. A slow drift in the position of our laser spot resulted in a roughly linearly decreasing detection efficiency as $\omega/2\pi$ was ramped from 8MHz to 9MHz. Figure S5 compares the signal in our "reference" region with and without renormalizing to correct for the slowly evolving detection efficiency. This renormalization is used only in computing $\omega_{1,\text{eff}}$ and $\gamma_{1,\text{eff}}$, and all other figures here and in the main text depict raw data.

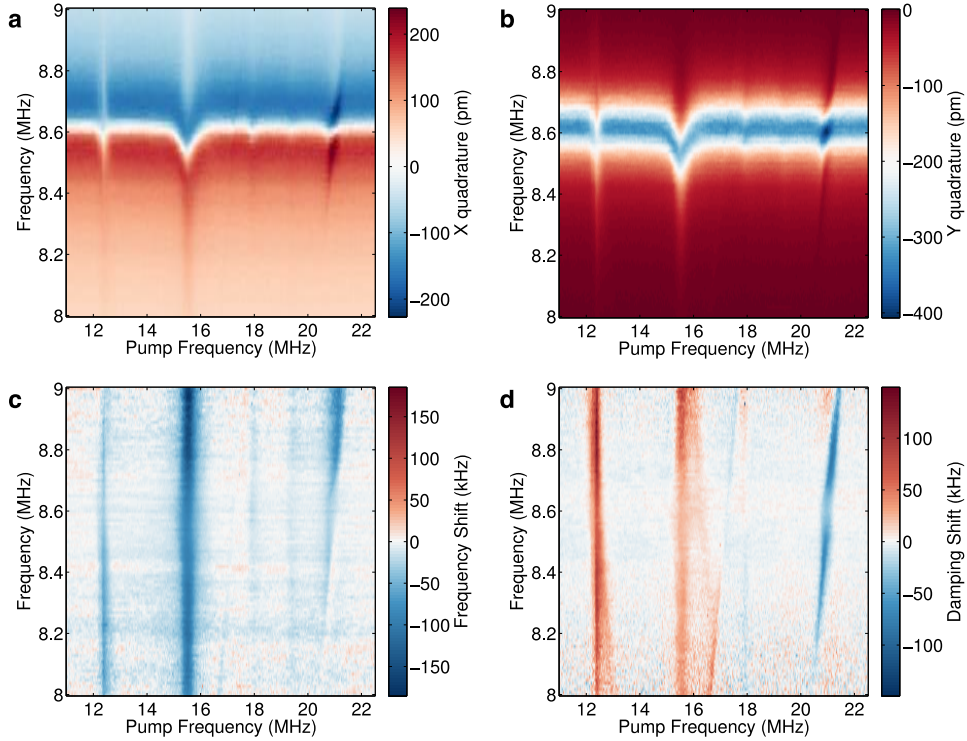

**Figure S4: a,b,** X & Y quadrature of Device 1 motion corresponding to Figure 3a of the main text. **c,d,** $\Delta\omega_1$ and $\Delta\gamma_1$ calculated from from X & Y according to Equations S49-S50. Note that these two are $\approx 0$ (by definition) in the "reference" region $\omega_p/2\pi = 22 - 22.5$ MHz, as well as most other regions. In these lower panels, the color scales are symmetric about 0kHz so that zero shift appears white. The intrinsic parameter values for mode 1 are $\omega_1/2\pi = 8.62$ MHz and $\gamma_1/2\pi = 150$ kHz.

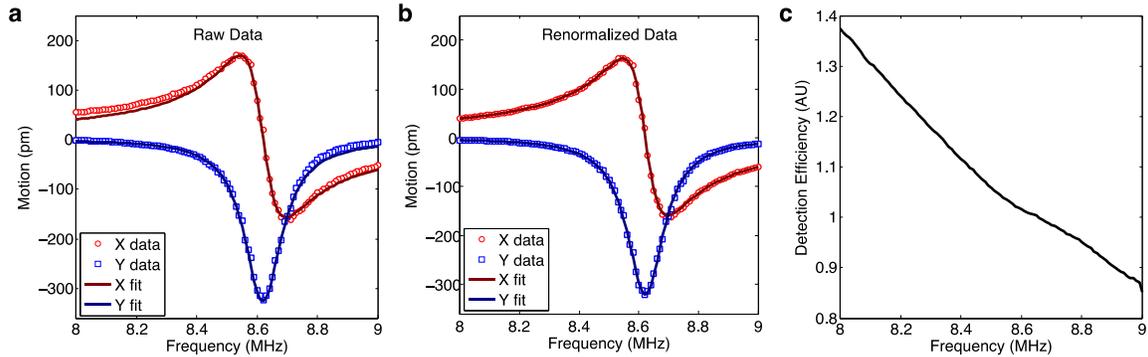

**Figure S5: a,** Raw data (X & Y quadratures) compared to a fit of Equation S48. Discrepancies are caused by a slowly decreasing detection efficiency over time (the frequency sweep shown was performed over ~1 hour). **b,** The same data, corrected for the changing detection efficiency. The near-perfect agreement between data and model is needed to ensure $\Delta\omega_1, \Delta\gamma_1 = 0$ in this "reference" region of $(\omega, \omega_p)$ space (see Figure S24. **c,** Detection efficiency used to renormalize the data and produce Figures S4c,d & 3e.

## S5 Additional mode coupling effects in Device 1

Measurements for Device 1 were taken at $V_{dc} = 2.5V, 5V, 7.5V, 10V$, and $15V$. As seen in Figure 2d of the main text, significant overlap between the frequencies of modes 2,6 and their respective sidebands occurs in the range $V_{dc} = 0 - 7.5V$. Therefore coupling rates between the modes should be roughly constant in this range, which is verified in Figure S6. Here care was taken to ensure equivalent drive forces ($F \propto V_{dc}V_{ac}$) were applied to the pump (as well as the probe) for each $V_{dc}$ value. At $V_{dc} = 10V$ and above, the sideband overlap diminishes and enhancement at the pump frequency no longer occurs.

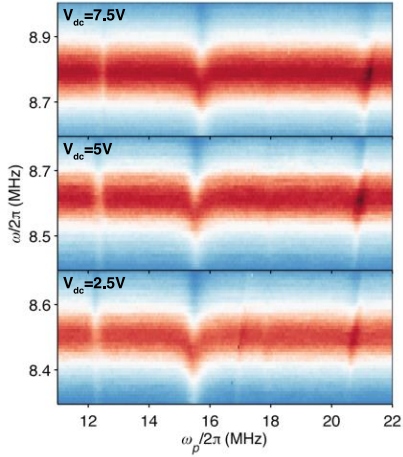

**Figure S6:** Mode coupling in Device 1 at 3 different $V_{dc}$ values. Drive forces applied at the pump frequency (and probe frequency) are equal across the three data sets. Apart from a steady increase in $\omega_1$ with $V_{dc}$, the coupling rates leading to amplification and cooling remain roughly constant.

As the graphene devices studied here have two independent back-gates, several driving conditions can be realized (depicted in Figure 1a). For the results shown in Figures 3 & S6 one back-gate is driven while the other is grounded. This configuration enables efficient actuation of all 6 membrane modes, as opposed to driving both back-gates in phase (which benefits the fundamental mode) or 180° out of phase (which benefits the higher modes). Figure S7 shows the results of driving the back-gates out of phase in Device 1 at $V_{dc} = 10V$. Interestingly, this strong driving of mode 4 (as well as its overlap with $2 \times \omega_1$) results in strong amplification of mode 1. This coupling between modes 1 and 4 is due to the $T_{14}x_1x_4^2$ term in the membrane Hamiltonian, Equation 1. Strong coupling between two mechanical modes $i, j$ where $\omega_j = 2 \times \omega_i$ has been studied previously in carbon nanotube systems[9].

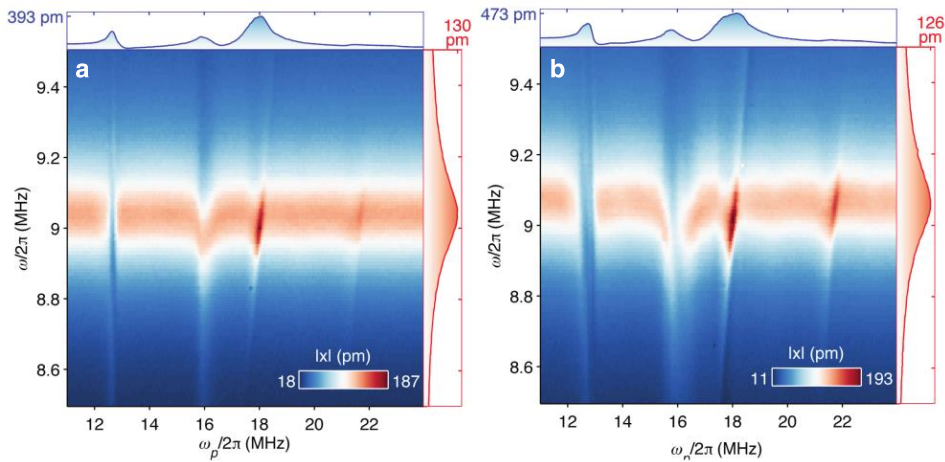

**Figure S7:** Mode coupling measuremetns with back-gates driven 180° out of phase. The pump voltage $v_{ac}$ is doubled between **a** and **b**. Both plots show parametric amplification at $\omega_p = 2 \times \omega_1$. For very strong pumping (**b**), there is also increased damping at $\omega_p = \omega_3 \approx 16MHz$. This is a separate effect from sideband cooling, and not yet fully understood. This feature is also seen in Figure S4d.

## S6  Large-amplitude quenching of phonon cavity and sideband mode

As discussed in the main text and seen in Figure 3, Device 1 exhibits quenching of the sideband amplification and cooling effects due to the large vibration amplitudes of mode 1. This is caused by nonzero couplings $T_{16}$ and $T_{61}$ which lead to effective cooling of either sideband mode due to motion of mode 1. The experimental data was modeled by (for the case of $\omega_c = \omega_2$, $\omega_{sb} = \omega_6$) using Equations 2 & 3 to solve for the effective frequency and damping of mode 1, $\omega_{1,\text{eff}}$ and $\gamma_{1,\text{eff}}$, while concurrently solving identical equations for $\omega_{6,\text{eff}}$ and $\gamma_{6,\text{eff}}$. In this case, mode 2 acts like a phonon cavity for mode 6, and mode 1 is pumping its red sideband $\omega_2 - \omega_6$; mode 6 is thus cooled while mode 1 is being amplified. If this process is started with a low enough mode 1 vibration amplitude, significant amplification can occur such that $\gamma_{1,\text{eff}} \to 0$ before the sideband mode experiences much cooling (as is the case in our measurements of Device 2).

Figure S8 demonstrates our method for fitting the measured $\gamma_{1,\text{eff}}$, and demonstrates predicted behavior with and without quenching.

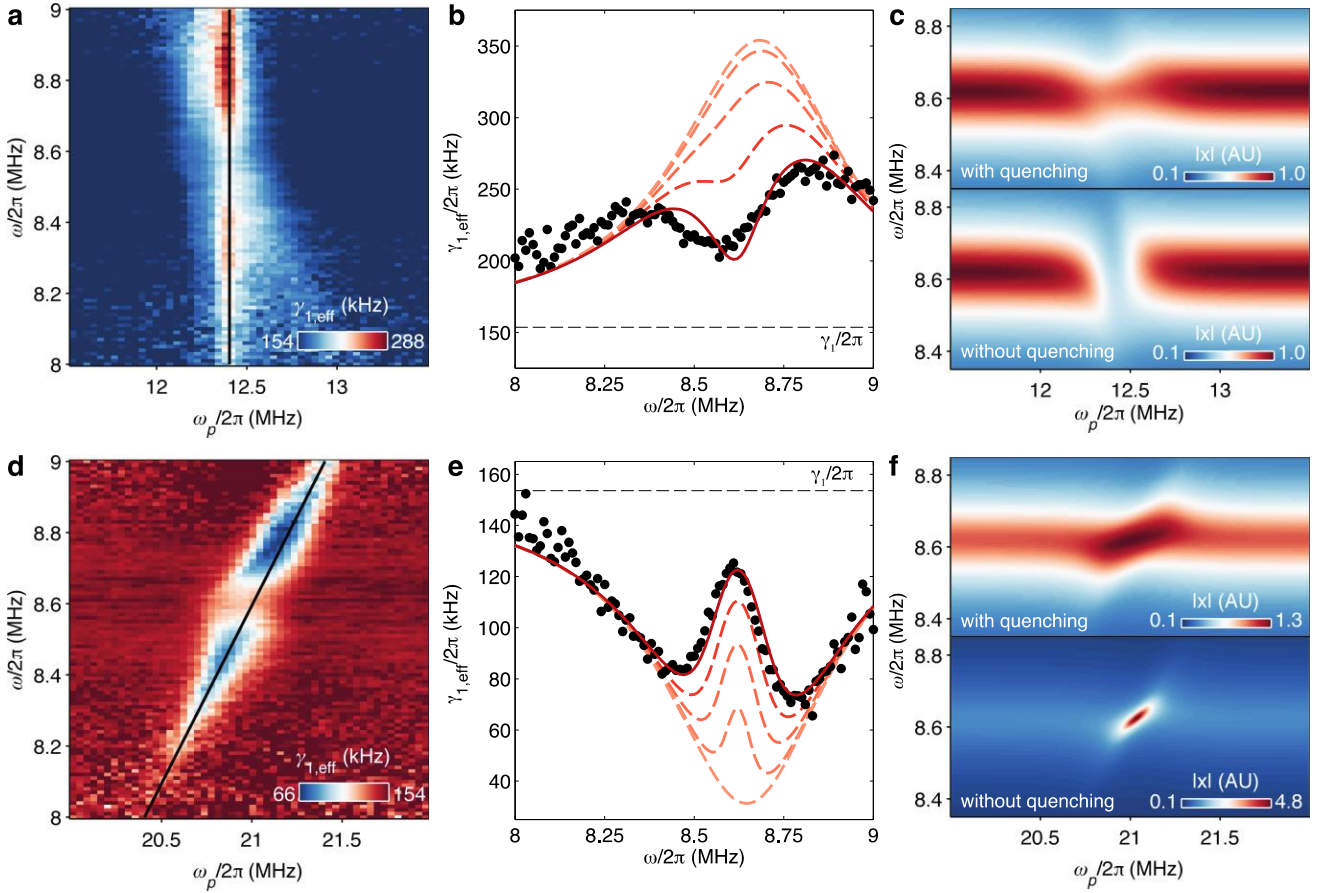

**Figure S8: Modeling the effective damping. a,** The measured mode 1 damping (obtained by the analysis described in Section S3) during sideband cooling. **b,** Fit to the data in **a**. Black points represent a slice through the data in **a** at the solid black line. The fit (solid red line) has two free parameters, $T_{16}$ and $T_{26}$, where the latter signifies quenching of mode 2. Dashed lines denote the same model for decreasing mode 1 amplitude (or equivalently, decreasing $T_{26}$) at 75%, 50%, 25%, and 0% of the experimental value. **c,** Simulated data with the fit parameters from **b**. Amplitudes are normalized to 1 when no cavity effects are present. **d-f,** Similar results for the sideband amplification effect.

## S7 Mode coupling in a third device

The parametric cooling and amplification effects described here and in the main text have been observed in several graphene membranes of various diameters. These effects are shown in Figure S9 for a third device (for which optical calibration has not been performed, and so motional amplitude is reported in photodetector uV).

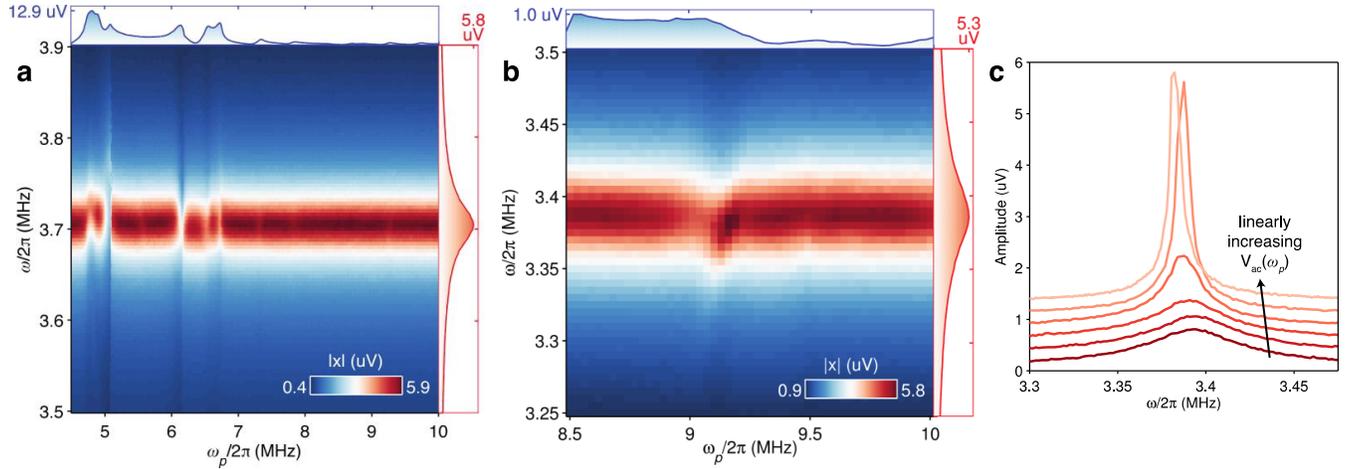

**Figure S9: Mode coupling in a 3$^{rd}$ device, diameter 16μm. a,** Mode coupling in this device at $V_{dc} = 10$V. Note the nontrivial spectrum of higher modes in the upper panel. Each mode coincides with increased damping of mode 1, suggesting sideband cooling via coupling to cavity modes in the $\omega_p/2\pi = 9-10$ MHz range. Modes in this range are not clearly visible (upper panel), possibly due to poor capacitive actuation to these modes. **b,** Mode coupling in the same device at $V_{dc} = 5$V. Some sideband amplification is visible. **c,** Amplification of mode 1 at $V_{dc} = 5$V upon pumping the sideband at $\omega_p/2\pi \approx$ 9.25MHz shown in **b**.

## S8 Duffing response & nonlinear damping

As suggested by Equation 1, motion-induced tension modulation results in other types of mechanical nonlinearity aside from inter-modal coupling. To demonstrate this, the response of Device 1, mode 1 was measured for strong drive amplitudes at $V_{dc}$=5V and $V_{dc}$=10V, as shown in Figure S10. Interestingly, for $V_{dc}$=5V, the resonance peak transitions from left-leaning at intermediate drive values to right-leaning at the highest drives. The intermediate amplitude behavior can be explained by a negative Duffing coefficient generated by $T_a$ and $F_a$ in Equation 1, while the transition to right-leaning is indicative of a higher order Duffing-like term (possibly $H \propto x_a^5$ or $x_a^6$). In Figure S10a and S10c, data has been modeled by a resonant frequency $\omega_{1,\text{eff}}^2 = \omega_1^2 + D_1|x_1|^2 + D_2|x_1|^3 + D_3|x_1|^4$, which reproduces the curved spine observed in the data of Figure S10a. Data in Figures S10d and S10f were modeled using only $\omega_{1,\text{eff}}^2 = \omega_1^2 + D|x_1|^2$.

Figure S10 also suggests that Device 1 undergoes nonlinear damping, as has been observed previously in graphene and carbon nanotube resonators[10]. The source of this nonlinear damping has thus far not been determined, and warrants further study. The fits shown in Figure S10c and S10f also include a nonlinear damping term $\gamma_{1,\text{eff}} = \gamma_1 + N|x_1|^2$, which reproduces the data well. However, the decreasing amplitude observed in Figures S10b and S10e may be partly due to a decreasing detection efficiency over time (the data was acquired over a span of 15 minutes), as described in section S4.

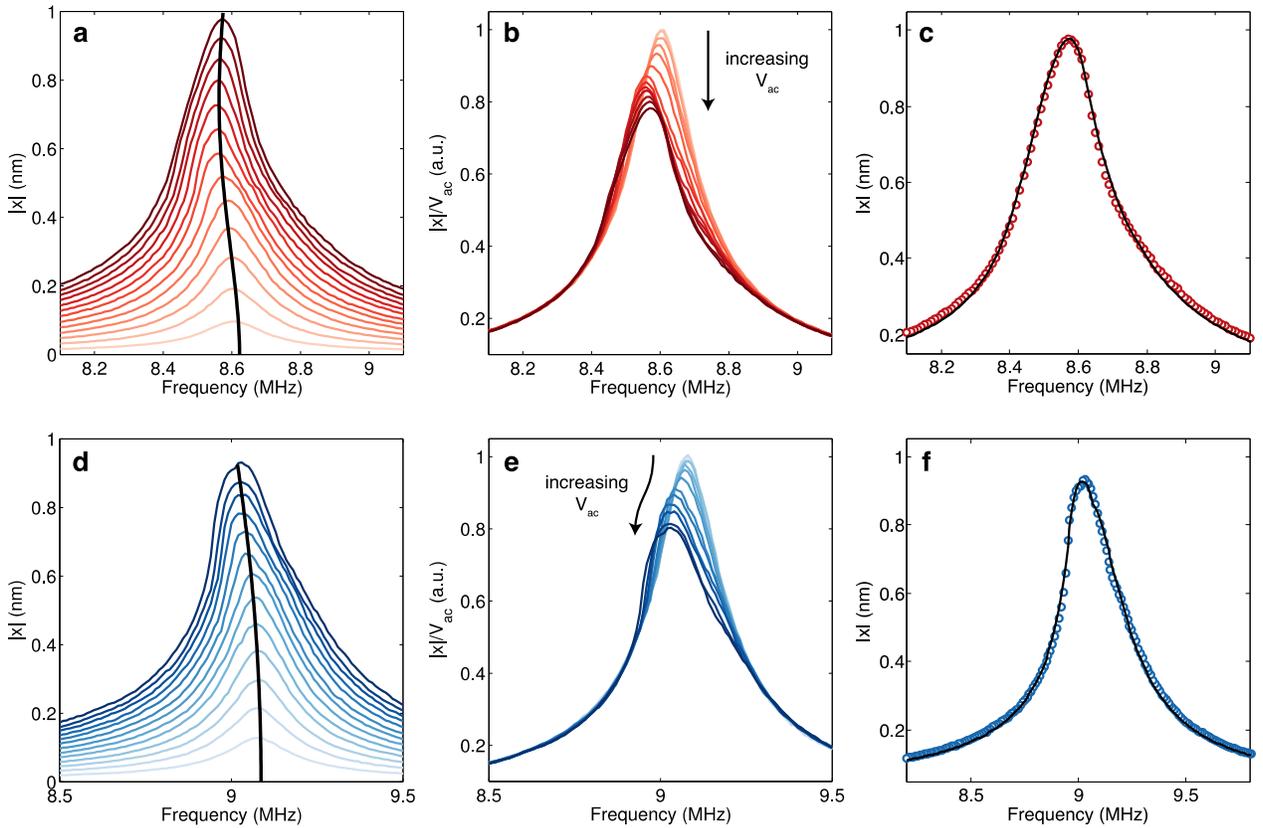

**Figure S10: Duffing response of Device 1. a,** Mode 1 response as drive amplitude is ramped from $v_{ac} = 4\text{mV}_{\text{rms}}$ to $56\text{mV}_{\text{rms}}$ (colored lines) at $V_{dc}$=5V. The black central line is a spine extracted from a fit to the highest curve (shown in **c**). **b,** The same data from **a**, normalized by ac drive voltage. The decreasing peak height is indicative of nonlinear damping. **c,** A fit the highest curve in **a**, with Duffing terms and nonlinear damping included. **d-f,** Similar data at $V_{dc}$=10V and $v_{ac} = 3\text{mV}_{\text{rms}}$ to $30\text{mV}_{\text{rms}}$.